# Electric field induced reversible control of visible photoluminescence from ZnO nanoparticles

Manoranjan Ghosh[†,#] and A. K. Raychaudhuri[$]

*DST Unit for Nanoscience, S. N. Bose National Centre for Basic Sciences,*

*Block-JD, Sector-3, Salt Lake, Kolkata-700 098, INDIA*

Reversible control of the photoluminescence of ZnO occurring in the visible range, has been achieved by application of a few volts (< 5V) to a device consisting of nanostructured ZnO film sandwiched between Indium Tin Oxide electrode and polyethylene oxide-lithium perchlorate, a solid polymer electrolyte. The photoluminescence intensity shows nearly 100% modulation with a response time less than 30 seconds, when the bias is applied at the electrolyte-electrode. A model is proposed for the observed effect that is based on defect states of ZnO and the band bending at the ZnO-electrolyte interface that can be changed by the applied bias.

[†]*Present address: Bhabha Atomic Research Centre, Mumbai – 400 085, INDIA*

[#]Email: mghosh@barc.gov.in , [$]E mail: arup@bose.res.in

ZnO, a wide band gap (3.3 eV) semiconductor with wurtzite structure (P63mc) is a well-known material for applications in the ultra violet (UV) region of the electromagnetic spectrum. Over the past decade, tremendous progress in research has been achieved for the use of ZnO as an optoelectronic material.[1] In particular, nanostructured ZnO that can be fabricated at temperatures below 100 $^0$C, can be used for innovative applications. The room temperature photoluminescence (PL) of ZnO nanoparticles (NPs) of size below 20 nm typically exhibits near band edge (NBE) emission in the UV region and a broad emission band in the visible region of the spectrum.[2] Fine control of this defect related visible PL is challenging because of its dependence on many environmental parameters. Factors like size and shape of the material, polar nature of the surrounding medium and type of defects involved in the emission processes were identified and the positively charged oxygen vacancies were commonly attributed as likely origin of this emission.[3,4] It will be an interesting proposition if a full reversible control of this visible emission is achieved by application of a voltage bias. This will not only be of scientific interest but can also have an application potential.

NBE emission of a semiconductor, that has an excitonic origin, can be controlled by application of electric field at the semiconductor-metal (S-M) and semiconductor-electrolyte (S-E) junction. For example, reversible change in excitonic PL from epitaxial GaN[5] was seen by application of electric field at the S-M junction. In a recent report, reduction of the visible emission and enhancement of the UV emission has been observed in a ZnO based Metal-Insulator-Semiconductor (MIS) structure on application of a bias.[6] Reduction of PL from n-GaAs electrode immersed in aqueous ditelluride electrolyte by application of external voltage has also been reported.[7] Recently it has been demonstrated by us that ZnO NPs suspended in an electrolytic solution show reduced visible emission.[8] In this paper we describe reversible control (reduction as well as enhancement) of the visible PL from ZnO NPs by application of small voltage (< 5 V) at S-E junction. The observed enhancement as well as reduction is substantial.

A device consisting of Indium Tin Oxide (ITO)/ZnO NPs/polymeric electrolyte (figure 1) has been constructed to demonstrate the effect field controlled visible PL. The film of ZnO NPs of size ~10 nm [see inset of Figure 2 (a)], synthesized by solution route[9] is obtained by adding NP-ethanol dispersion on ITO coated glass substrate at room temperature. To ensure the presence of a ZnO layer all over the surface, a thick film is deposited. The average thickness was found to be ≈ 200-300 nm. When the ZnO film dries off, a gel of polyethylene oxide (PEO, MW 100000) and lithium perchlorate (*LiClO*$_4$) is deposited on the ZnO layer. After the gel becomes solid, it holds securely the electrical connections (Cu wires) as shown in Fig. 1. The PL spectra were measured by a Spectrofluorimeter (Jobin Yvon Fluromax 3). The bias voltages were supplied by a Stanford Research System DS345 synthesized function generator.

The device has been illuminated by UV light of energy higher than the band gap of ZnO. However, the upper energy is limited by the transmittance of ITO. We excite the device through ITO glass face and the emitted signal is collected in the reflection mode. In this arrangement, the thick ZnO film acts as an absorbing layer before the illumination reaches the electrolyte at the interface. This eliminates the PL from the polymer that occurs at around 425 nm. ZnO NPs sandwiched in the device show maximum NBE PL near 390 nm and visible PL around 500-550 nm when illuminated in the window 320-355 nm through ITO. (See supplementary material at [URL] for absorption and PL spectra of individual component). The PL spectra from the device [Fig. 2 (a)], matches with that obtained from the colloidal ZnO NPs taken in suspension.[3,8] However, the relative intensity of the surface related broad visible PL is reduced due to the surface passivation of ZnO NPs by *PEO-LiClO$_4$* layer. A shoulder around 400-435 nm superimposed on the NBE emission appears in the PL spectra due to the emission from PEO-LiClO$_4$ as stated before. However, this does not affect the visible PL signal near 500- 550 nm range.

The main result of this work is the remarkable change in the visible PL from ZnO NPs by application of voltage bias. Nearly 100 percent modulation of the visible PL intensity is achieved [Fig. 2(a)] while maximum change occurs within ± 1.5 V. PL intensity increases when the device is biased as depicted in figure 1 and vice versa. In Fig 2(b), we show voltage dependence of the ratio $I_{VIS}/I_{NBE}$, where $I_{VIS}$ and $I_{NBE}$ are the intensities of visible (~550 nm) and NBE (~390 nm) emissions respectively. $I_{VIS}/I_{NBE}$ continuously decreases when the bias is changed from -2 V to +2 V. The ratio can be changed by nearly a factor of 3 when the bias is cycled between ±3V and it saturates for bias beyond that. The bias cycling at a finite rate can cause a hysteresis effect in the observed values of $I_{VIS}/I_{NBE}$ due to finite response time. For the data presented in Fig 2(b), the scan time for the full voltage cycle is ≈ 30 min. For longer cycling time the hysteresis becomes smaller and eventually vanishes.

The finite response time seen in the device arises due to polymeric nature of the electrode. To quantify the response time, the PL intensity is measured in response to a square wave voltage pulse of amplitude ± 4.5 V at frequencies 0.001 Hz [figure 3 (a)]. In addition, an amplitude modulated (varying from 0 - 4.5 V) square wave of frequency 0.01

Hz has been applied to demonstrate precise control over the intensity by voltage [figure 3 (b)]. The visible PL exactly follows the amplitude and polarity of the applied voltage as shown in figure 3. Up to 105 % enhancement and 85% reduction is seen for negative and positive half cycles respectively. PL intensity initially shows faster voltage response and then relaxes following an exponential time dependence to a stable value with a long relaxation time (≈ 20-30 sec).

The visible emission from ZnO in the blue-green region originates from defects (oxygen vacancy in particular) which are located predominantly near the surface of the NPs.[3,4] It is also seen that this emission is a composite of two broad lines located approximately at 2.2 eV (550 nm) and 2.5 eV (500 nm). Emission band appearing around 550 nm (which we call $P2$) has been suggested to originate from doubly charged oxygen vacancy ($V_o^{++}$) which is predominant in ZnO nanospheres. Whereas singly charged oxygen vacancy ($V_o^+$) is responsible for the emission band around 500 nm (which we call $P1$).[3,4,8] It has been shown that emission intensity (comprising the intensities of the two lines) depends on the ionic environment and the zeta potential of the NPs[8]. Emission intensity reduces when the zeta potential of the NPs decreases with larger change in the P2 line. The effect was explained in terms of band bending near the surface of the NP that depends on the population of the emitting species like charged oxygen vacancies and the width of the depletion layer. Earlier reports, including the investigations done by our group establish that the emission intensity from the NPs can be controlled by changing the band bending near the surface.[3,4,8]

The observed phenomena, presented in this paper, can be explained using similar arguments involving band bending. The band bending and hence the population of the positively charged oxygen vacancies near the surface of the NPs can be changed at the ZnO-electrolyte interface by application of a bias. When ZnO is immersed in the electrolyte such as PEO-LiClO$_4$, the interface will be governed by the alignment of the redox potential ($E_{redox}$) of the electrolyte with the electrochemical potential (Fermi level $E_F$) of the semiconductor.[10] When they are different, this will lead to carrier transfer across the interface which causes band bending. In the present case, when the n-type ZnO NPs are made to touch the electrolyte, there will be transfer of electrons from the ZnO to the electrolyte. Depletion of the majority carrier will create a depletion region in ZnO and the band bending occurs. The schematic of the band positions for the n-ZnO and the electrolyte in equilibrium (without bias) is shown in figure 4(a).

When a positive voltage is applied to the electrolyte, the electrolyte side of the ZnO-electrolyte interface accumulates the $Li^+$ ions which induces more negative charge in the surface region of ZnO NPs [figure 4 (b)].[10] The accumulation of the negative charge fills up the positively charged oxygen vacancies. This reduces the visible PL. The presence of more majority carrier reduces the band bending and can even lead to creation of accumulation layer. The presence of majority carriers in the accumulation layer also shows up as a substantial enhancement of the junction current. A reverse situation arises when negative bias is applied to the electrolyte[10] as depicted in figure 4(c). It induces more positive charge on the ZnO surface. This increases the number of ionized oxygen vacancies and leads to enhancement of the PL. In this situation the depletion width ($W_D$) increases due to depletion of the majority carriers and the upward band bending becomes more severe. The presence of the depletion layer leads to reduction of the junction current by at least one order compared to that in the accumulation case.

Use of the polymer electrolyte is essential to achieve the reversible control over the visible PL. To test this hypothesis, we make a metal-semiconductor device consisting of ZnO NPs and a top Al layer. This exhibits small enhancement in NBE emission and a small reduction in visible PL at a much higher positive voltage (~ 12 V) applied to the Al (data not shown). No enhancement in visible PL is seen. It is likely that the polymer electrode produces electric double layer at the interface and allows reversible control of band bending at lower voltages.

To summarize, a reversible control of the visible PL is demonstrated in films made from ZnO NPs, by the application of an electric field in a simple device geometry that uses polymer electrolyte as an electrode. Depending on the sign of the voltage bias, the emission intensity can be enhanced or reduced. It is proposed that the application of voltage bias controls alignment of the $E_F$ and $E_{redox}$. This affects the band bending at the interface and filling of the $V_o^+$ and $V_o^{++}$ states that control the emission intensity. The voltage control (by a small bias <5V) that can enhance as well quench the visible emission has an application potential that is being explored.

The authors would like to thank Department of Science and Technology, India for the financial support as a unit for Nanoscience.

[1]U. Ozgur, Y. I. Alivov, C. Liu, A. Teke, M. A. Reshchikov, S. Dogan, V. Avrutin, S-J Cho and H. Morko, J. Appl. Phys. **98**, 041301 (2005).

[2] B. K. Meyer, H. Alves, D. M. Hofmann, W. Kriegseis, D. Forster, F. Bertram, J. Christen, A. Hoffmann, M. Straßburg, M. Dworzak, U. Haboeck, and A. V. Rodina, phys. stat. sol. (b) **241,** 231 (2004).

[3]M. Ghosh and A. K. Raychaudhuri, Nanotechnology **19,** 445704 (2008).

[4] K. Vanheusden, C. H. Seager, W. L.Warren, D. R. Tallant and J. A. Voigt, Appl. Phys.Lett. **68**, 403 (1996).

[5]D. K. Nelson, V.D. Kagan, E.V. Kalinina and M.A. Jacobson, Journal of Luminescence **72-74**, 865 (1997).

[6]X. Ma, P. Chen, D. Li, Y. Zhang, and D. Yang, Appl. Phys. Lett. **91**, 021105 (2007).

[7]W. S. Hobson and A. B. Ellis, J. App. Phys. **54**, 5956 (1983).

[8]M. Ghosh and A. K. Raychaudhuri, Appl. Phys. Lett. **93**, 123113 (2008).

[9]M. Ghosh and A K Raychaudhuri, J. Appl. Phys. **100**, 034315 (2006).

[10]Adrian W. Bott, Current Separations **17,** 87 (1998).

FIG. 1. (Color online) Illustration of the device fabricated. The biasing arrangement demonstrated here (negative bias to electrolyte) leads to enhancement of intensity and vice versa.

FIG. 2. (Color online) (a) Voltage controlled PL spectra of the device consisting of spherical ZnO nanoparticles of dia~10 nm shown in the inset. (b) Intensity ratio of the visible emission to the near band edge emission $I_{VIS}/I_{NBE}$ as a function of the applied bias. The arrows indicate the direction of voltage cycling.

FIG. 3. (Color online) (a) Variation of the visible emission (at 544 nm) intensity when a square wave voltage of frequency 0.001 Hz and amplitude 4.5 V is applied. (b) Intensity profile when the amplitude of the square wave of frequency 0.01 Hz is varied from 0 to 4.5 V.

FIG. 4. Band diagram at the nanoparticle–electrolyte interface: (a) in equilibrium after the contact is made (no applied bias). Figure (b) and (c) depicts the situation when electrolyte is positively and negatively biased respectively.

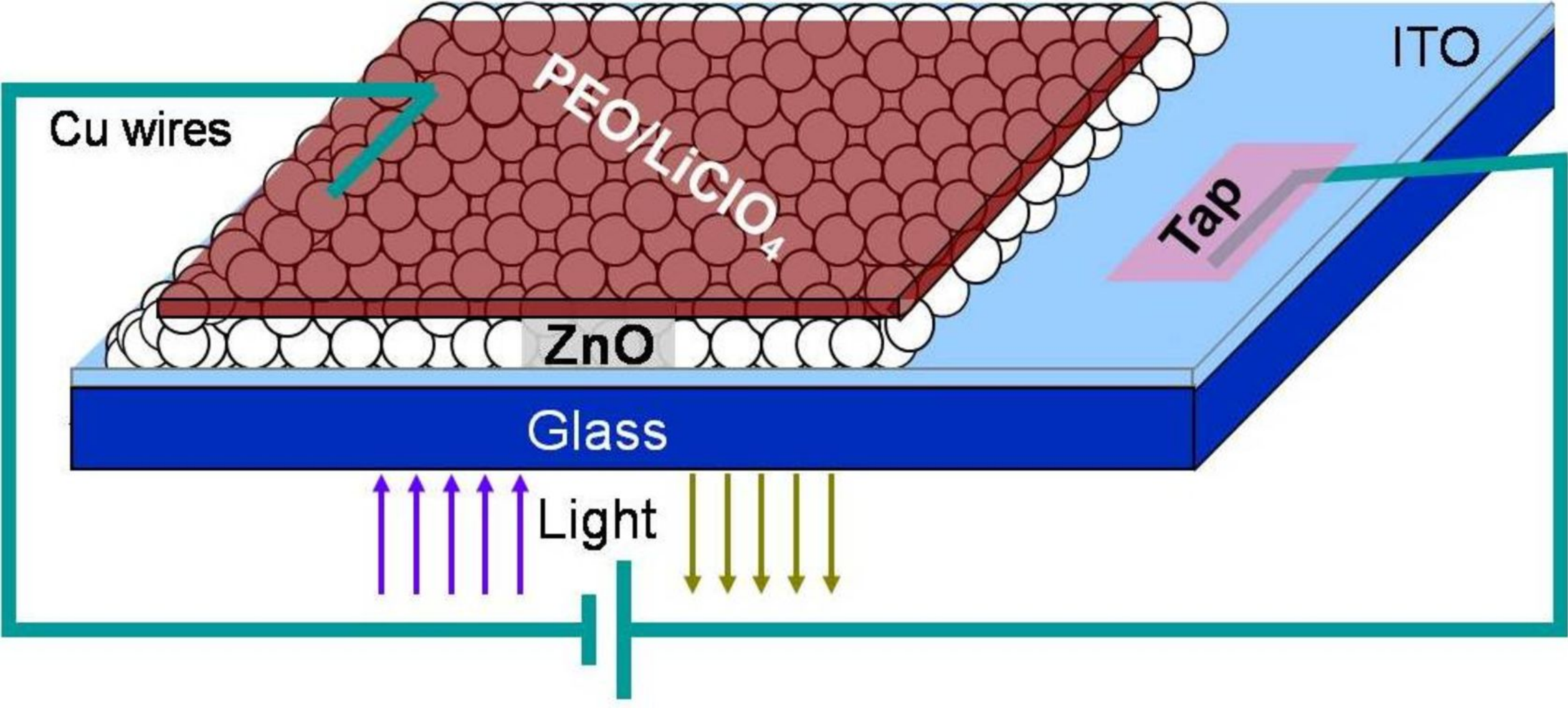

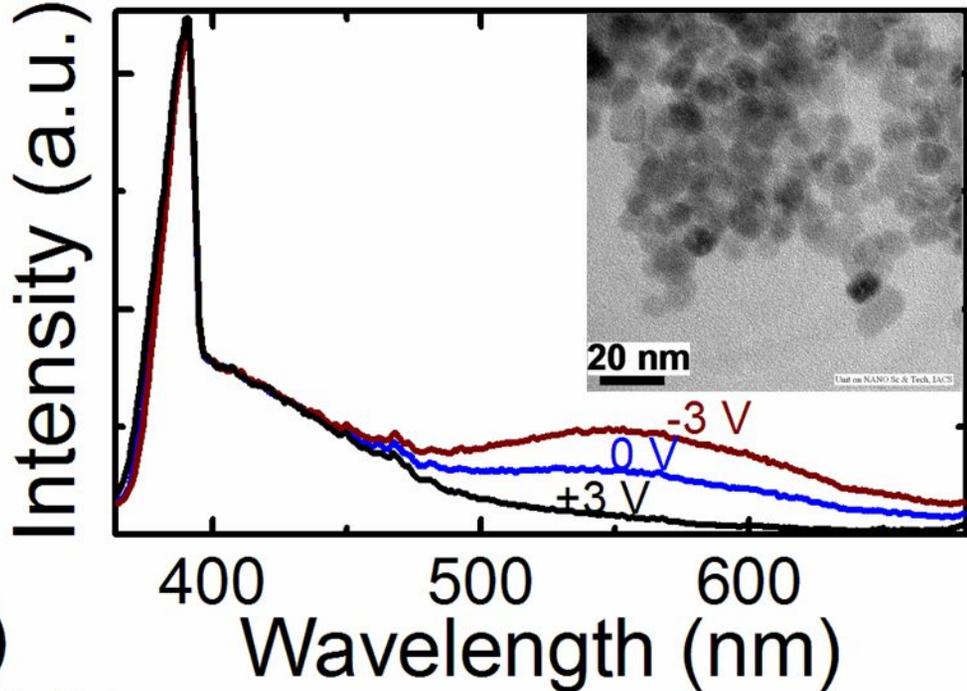

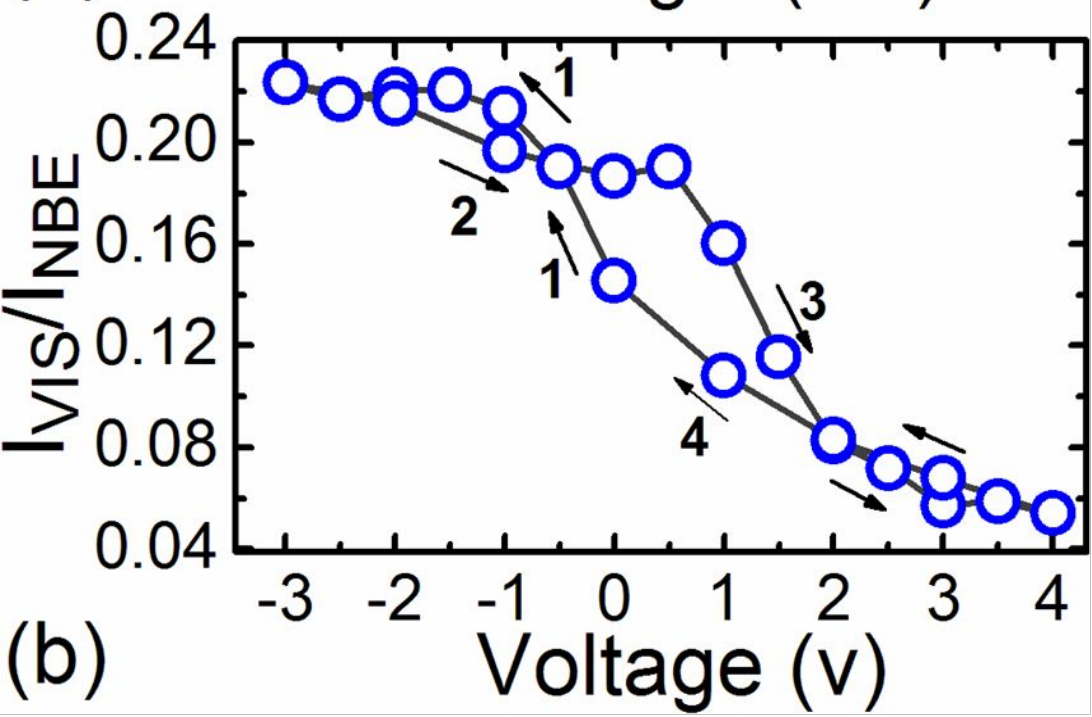

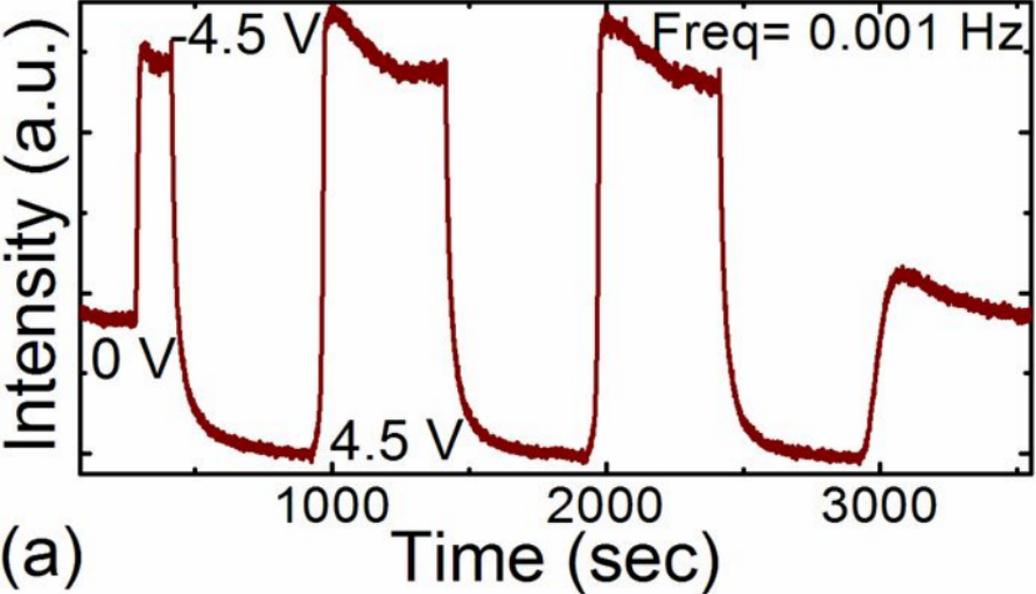

(a)

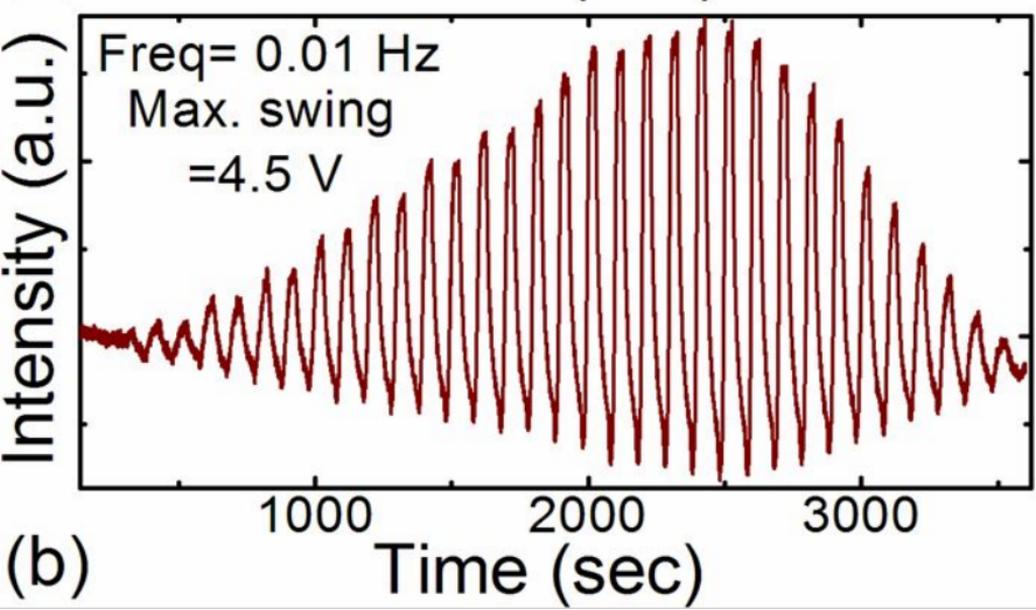

(b)

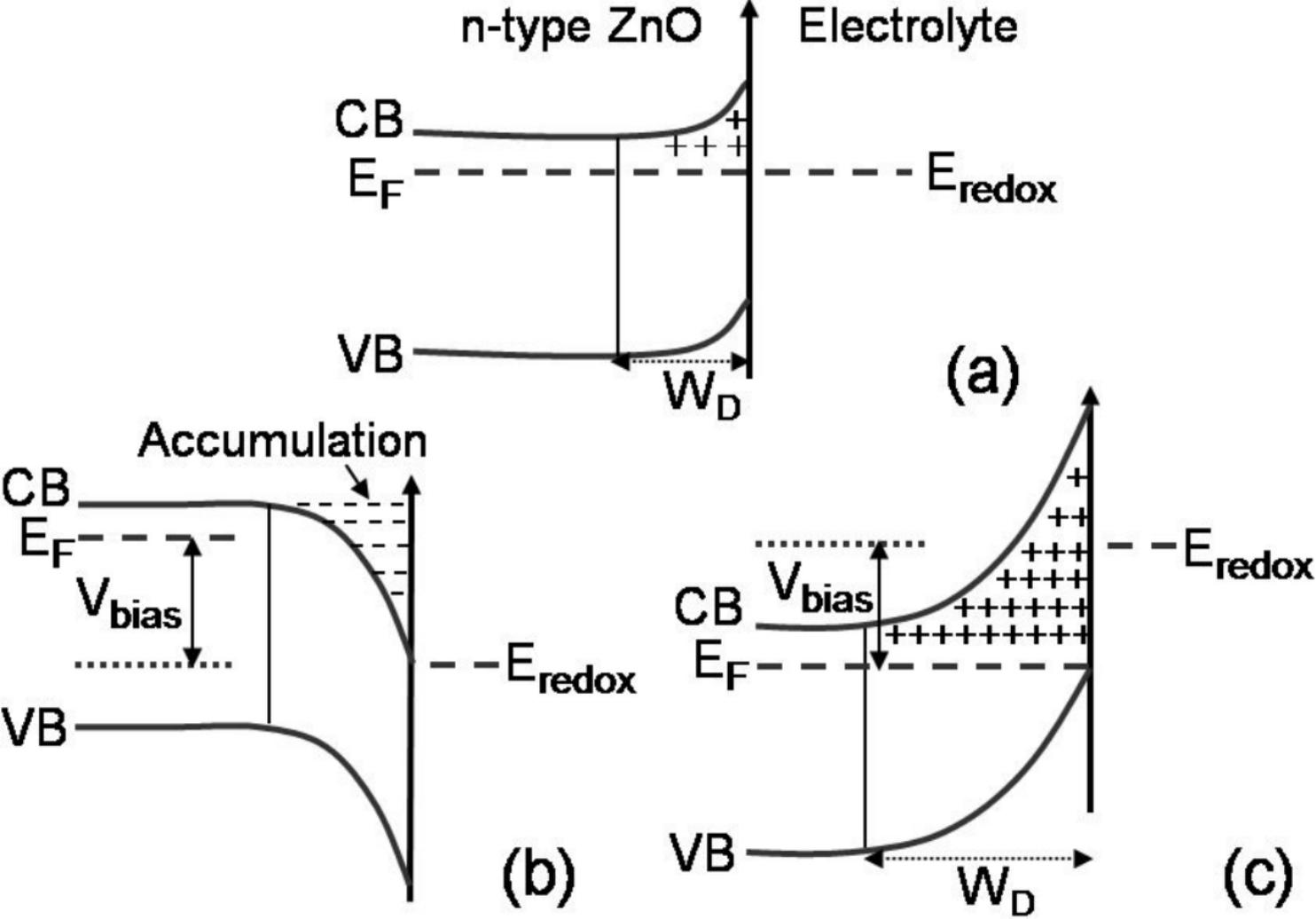